# Optimal Order of Multi-Agent and General Many-Body Systems

Jake J. Xia[1]

## Abstract

This paper develops a general framework for analyzing multi-agent systems with feedback loops between agents' actions and collective observations. The framework is built on two fundamental agent-level variables: power, which measures an agent's influence on collective outcomes, and response functions, which determine how agents react to observations. We derive how macroscopic properties—including total power, useful power, entropy, order, fragility, and mobility—emerge from these two variables of heterogeneous agents. To study the tradeoff between growth and resilience, we introduce a system-level utility function parameterized by a risk-appetite coefficient and derive an optimal degree of order that balances productivity, stability, and adaptability. The analysis suggests that stronger synchronization can increase collective output but may also increase systemic fragility and reduce mobility. We further argue that order, entropy, information, and useful energy are task-dependent and system-relative concepts whose meanings depend on the objectives of the system. By measuring and designing agent power distributions and response functions, it may be possible to better understand, predict, and optimize collective behavior and identify the conditions under which collective intelligence and optimal order emerge.

---

[1] Harvard Management Company, Research Affiliate at Massachusetts Institute of Technology. Email: jjxia@mit.edu. This paper is a revised version of a 2013 working paper [21].

# I. Introduction

## Background: what is a many-body system?

A many-body system is a collection of a large number of interacting elements (such as particles, agents, or neurons) whose collective behavior creates complex, emergent properties that cannot be understood by analyzing the isolated elements individually. Because of the vast number of interactions, exact solutions are typically impossible to calculate, requiring advanced approximations, statistical mechanics, and specialized mathematical frameworks. Purely reductionist approaches are often insufficient for emergent many-body behavior. One classical example is thermodynamic system where statistical mechanics is applied.

Multi-agent system is a subset of many-body systems, usually referring to self-organized systems composed of multiple interacting intelligent agents. Social systems such as financial markets, economies, and social networks are typical multi-agent systems. Brain neural network can also be viewed as a multi-agent system where neurons act like agents with interactions. With recent advancements in large language models (LLMs) [Vaswani et al 2017], LLM-based multi-agent systems have emerged as a new area of research, enabling more sophisticated interactions and coordination among agents. Typically, research on multi-agent systems refers to software agents. However, the agents in a multi-agent system could equally well be robots, drones, vehicles, and humans.

Social systems differ from natural systems because agents can proactively respond and act based on a feedback loop. Power-law distribution is often observed in social systems compared with normal distribution in natural systems. In nature, typical many-body systems do not have feedback-loop interactions between actions of elements and observations. Multi-agent systems consist of self-organized autonomous agents that can respond to observations to alter collective outcomes. Some passive physical systems can be treated as limiting cases with negligible adaptive response functions. It is difficult or impossible to track and model each individual agent or a monolithic system. However, we can derive system-level macroscopic properties from agent-level behavioral patterns [Xia 2016].

In social systems of multi-agents, we often face the problem of optimization. How much concentration and synchronization of agents' power are optimal? Why market sentiment swings between risk-on and risk-off? What is the trade-off between economic growth and avoiding recession? What determines the outcome of group behavior between crowd-wisdom and crowd-madness? How much should government intervene from top down vs let agents act from bottom up? All these questions lead to, what is the optimal order? Is such optimality absolute (objective) or relative (subjective)?

Applications of multi-agent systems include crowd dynamics, markets, social networks, flocking/swarming, opinion dynamics, collective intelligence, epidemic behavior, artificial intelligence (AI) reinforcement learning models, AI multi-agent systems, brain neural systems, evacuation simulations, traffic congestion, reflexivity, and adaptive learning.

## Literature Review

The study of multi-agent systems (MAS) spans several disciplines, including game theory, economics, computer science, artificial intelligence, physics, and complex systems research. A common objective across these fields is to understand how interactions among many agents generate collective outcomes at the system level. Existing approaches include mean-field games, stochastic games, dynamical systems theory, and agent-based modeling. While each framework captures important aspects of collective behavior, none explicitly focuses on the role of feedback loops in determining the growth, useful power, stability, and vulnerability of a system. The framework proposed in this paper seeks to integrate elements of these approaches into a unified theory of feedback-driven social systems [Sun 2006].

### *Mean-Field Games and Stochastic Games*

Mean-field game (MFG) theory studies strategic decision-making in systems with a very large number of interacting agents. By approximating the influence of all other agents through a representative mean field, MFGs make otherwise intractable multi-agent optimization problems analytically manageable. The framework has been widely applied in economics, finance, engineering, and multi-agent learning.

A closely related framework is the Markov game, or stochastic game [Farina 2024], originally introduced by Shapley [Shapley 1953]. In a Markov game, the actions of multiple agents jointly determine probabilistic transitions between system states, while each agent receives rewards based on the current state and actions taken. A central solution concept is the Markov Perfect Equilibrium (also called Nash Equilibrium or Markov Stationary Nash Equilibrium) in which each player's strategy depends only on the current state and not on the historical path that generated it. Markov games have become a foundational framework for reinforcement learning and multi-agent artificial intelligence.

Our approach shares several features with stochastic games, particularly the dependence of system evolution on agents' actions and observations. However, unlike traditional game-theoretic models that emphasize strategic equilibrium, the present framework focuses on non-zero-sum systems in which all agents may simultaneously experience gains or losses. Furthermore, we emphasize deterministic feedback mechanisms that drive state transitions, treating stochasticity as a secondary component rather than the primary source of dynamics. By identifying the causal factors governing system evolution, the framework seeks to explain the origins of order, synchronization, and systemic instability.

### *Dynamical Systems Theory*

Dynamical systems theory provides a mathematical framework for studying how system states evolve over time. Such models are widely used in physics, engineering, biology, and economics to describe processes ranging from population growth to opinion formation. A dynamical system consists of a set of state variables and a rule governing their evolution, often expressed through differential equations.

When applied to social systems, dynamical models capture how local interactions aggregate into macroscopic outcomes. These systems may exhibit equilibrium, nonlinear behavior, instability, bifurcations, and chaos. In the linear case, rate of change of the state $x$ is proportional to the state itself: $dx/dt=\mathrm{k}x$. At equilibrium, $dx/dt=0$. Kauffman's work [Kauffman 1993] on self-organization demonstrated how simple dynamical rules can generate complex patterns of order. The framework developed in this paper can be viewed as a class of dynamical systems in which observations evolve through feedback interactions among agents. Linear formulations emerge as special cases, while more general nonlinear forms can capture richer collective dynamics.

### *Agent-Based Modeling*

Agent-based modeling (ABM) adopts a bottom-up approach to the study of complex systems [Epstein & Axtell 1996]. Rather than relying on representative agents or aggregate equations, ABMs simulate the behavior of heterogeneous agents interacting according to specified behavioral rules. Through these interactions, macro-level patterns emerge endogenously from micro-level decisions. Monte Carlo methods are used to understand the stochasticity of these models.

ABMs are particularly useful for studying systems characterized by feedback loops, heterogeneity, network effects, and nonlinear interactions. They have been applied extensively in economics, sociology, epidemiology, ecology, and organizational research. Recent applications include large-scale simulations of epidemic transmission, where agent-based models have often outperformed traditional compartmental models in capturing heterogeneous behavior and policy interventions.

A key advantage of ABMs is their ability to conduct counterfactual experiments and policy analysis that would be impractical or impossible in real-world settings. However, while ABMs provide detailed simulations, they often lack general analytical results. The framework proposed in this paper seeks to complement ABMs by providing theoretical measures of system-level properties such as total power, useful power, order, synchronization, stability, and vulnerability.

### *Multi-Agent Systems and Collective Intelligence*

A multi-agent system consists of multiple interacting autonomous agents that collectively perform tasks or solve problems that exceed the capabilities of individual agents. Research in MAS draws upon reinforcement learning, distributed optimization, complex systems, and collective intelligence. Modern applications include autonomous robotics, distributed computing, blockchain networks, social platforms, and large-scale AI systems.

This line of research highlights the broader relationship between coordination, robustness, and systemic stability, themes that are central to the present study.

### *Control Theory, Bayesian Learning, and Feedback Systems*

Control theory, Bayesian learning, and feedback systems provide the mathematical foundations for understanding prediction, adaptation, and stability in dynamic environments. These fields study how a system observes its environment, updates its internal state, and adjusts its actions in response to new information. Current observations influence future actions, and those actions subsequently alter future observations. Classical control theory focuses on the regulation (stabilization) of dynamic systems through feedback. Adaptive control extends these ideas to environments in which system parameters themselves may change over time.

Bayesian learning provides a complementary framework for decision-making under uncertainty. Bayesian methods treat the true state of a system as partially observable and continuously update beliefs as new information becomes available. The Bayes filter formalizes this process through a recursive cycle consisting of prediction and updating. This recursive process enables agents to learn from experience while accounting for uncertainty and measurement noise. The Kalman filter represents the canonical solution to Bayesian state estimation in linear Gaussian systems.

The framework developed in this paper places response functions at the center of analysis. Rather than focusing solely on state estimation or optimal control, we model social systems as networks of heterogeneous agents connected through feedback loops between observations and actions. From this perspective, control theory studies how a system should respond to observations, Bayesian learning studies how observations modify beliefs, and our framework studies how the response functions of many interacting agents jointly generate the evolution of the system itself.

### Focus of This Paper

The framework proposed in this paper integrates insights from stochastic games, dynamical systems, and agent-based models while emphasizing feedback-loop interactions as the fundamental mechanism governing collective behavior. Rather than focusing solely on equilibrium outcomes, state trajectories, or computational simulations, the framework seeks to derive macroscopic system properties—including total power, useful power, entropy, order, synchronization, stability, and vulnerability—from the interactions among heterogeneous agents.

In my 2016 paper [Xia 2016], crowd synchronization is modeled by agents' response functions to collective actions. I also applied this model to explain how power-law wealth distribution can emerge from interaction of agents [Xia 2024]. In this paper, I will further derive other macroscopic properties of a system from the same feedback-loop interaction between agents and observations. The core topic of this paper is how to determine an optimal order of a complex system. I will use utility functions and other objective functions to study this.

## II. General Model of Multi-Agent Systems

This section derives macroscopic properties of a many-body system from agent level interactions. Let's first review the model.

**Feedback Loop Model**

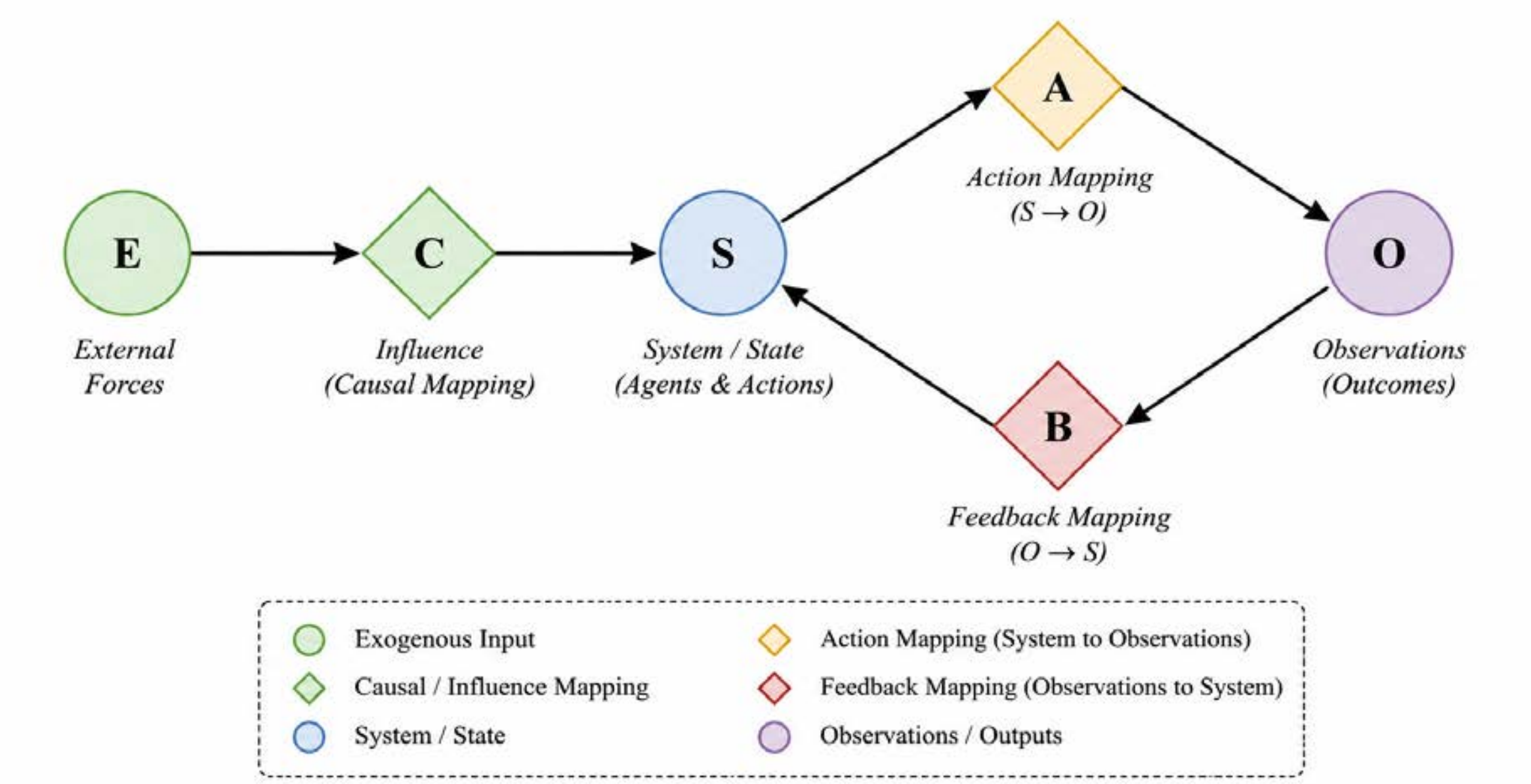

Figure 1a. Feedback loop illustrated by external force E with influence C, agent with action S and power A, observation O with feedback B.

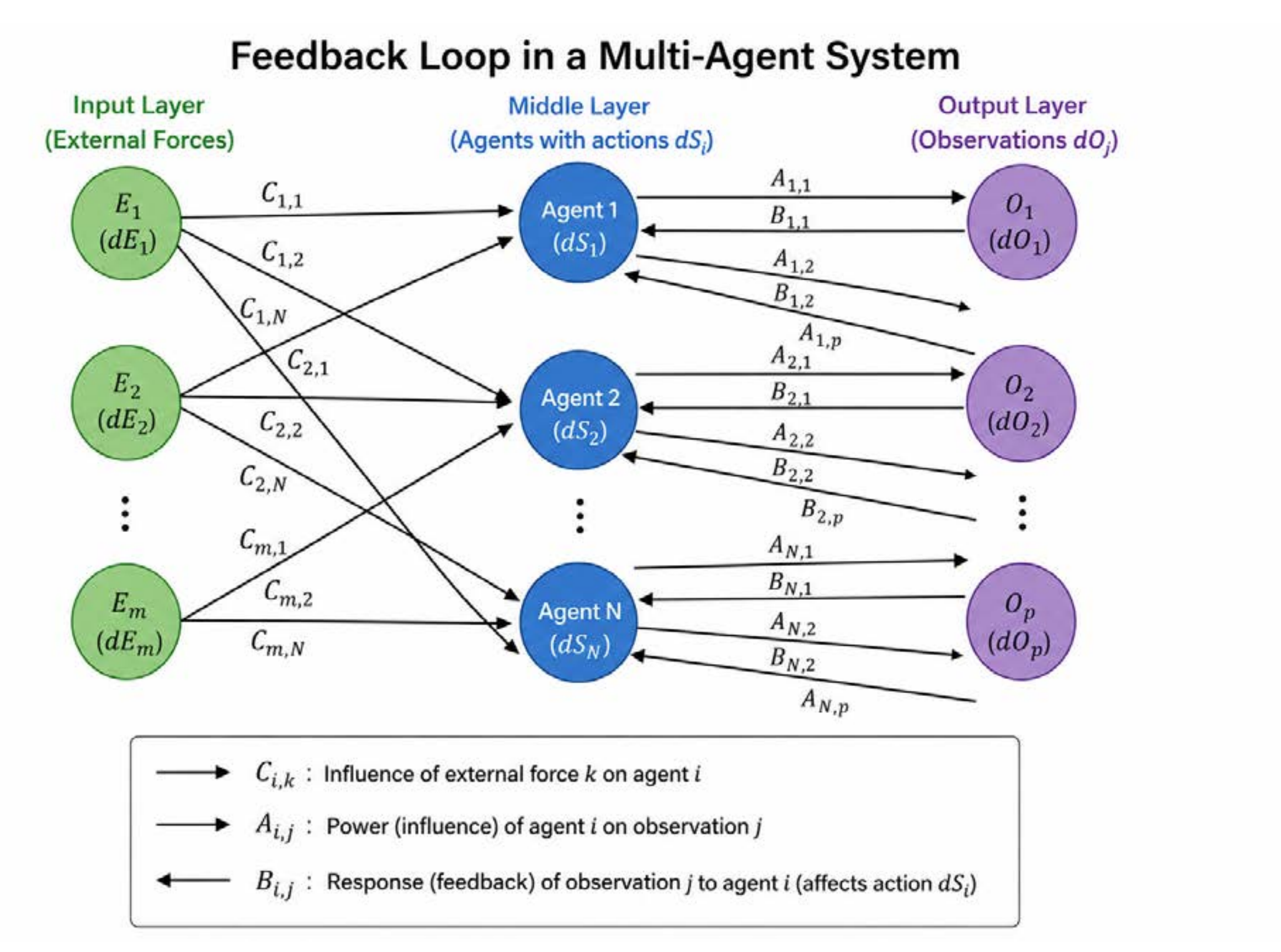

Figure 1b. Multi-agent systems with multiple observations $O_j$ (tasks, games or outputs) and feedback loops. Given limited space of the diagram, the arrows are only illustrative.

Instead of modeling local interaction between agents, we model the interaction between agents and observations. If only two agents are participating in a particular observation, the interaction is reduced to between the two agents. This approach is more general than modeling interactions between any two agents, because for different games/tasks/observations, agent's response functions may be different. The observations include environment as described in multi-agent systems in LLMs.

In this model, interaction of agents, observation, and external force is modeled as a feedback loop. Here, I also generalize the model to multi-observation or multi-task problems. In observation or task $j$, at each time step, agent's incremental change of decision $dS_{ij}$ is determined by its responses to external force relative change $dE_j$ (unitless) and observation change $dO_j$ plus some random noise $\varepsilon_{ij}$. $dO_j$ should be viewed as relative change (unitless) in observation parameter, more precisely $dO_j/O_j$ . For notation simplicity, we will continue to use $dO_j$ or often drop the subscript $j$ unless we are discussing multi-task situations.

$$dS_{ij}(t) = C_{ij}\, dE_j(t) + B_{ij}\, dO_j(t - dt) + \varepsilon_{ij}(t) \qquad (1)$$

Where *t* and *t-dt* are time steps, $C_{ij}$ is agent *i* 's response coefficient to $dE_j$, $B_{ij}$ is agent *i* 's response coefficient to $dO_j$ at previous time step and it is unitless, $A_{ij}$ is agent $i$'s energy or power or wealth or influence (we use these terms interchangeably) in game or task or observation $j$. $dS_{ij}$ is also unitless. The response functions don't have to be linear. For nonlinear cases, the response functions $B_{ij}$ and $C_{ij}$ are not simply coefficients.

In equation (1), $\varepsilon_{ij}(t)$ is a stochastic term including random noise of Wiener process and drift, describing agent $i$'s decision not determined by $dE_j$ or $dO_j$. For simplicity, in later discussion, we often assume the impact of $\varepsilon_{ij}$ is small compared with the first two terms and drop $\varepsilon_{ij}$ in modified equation (2).

$$dS_i = C_i\, dE + B_i\, dO \qquad (2)$$

The term $\varepsilon_{ij}$ is important when correlation between agents is considered, e.g. for calculating order parameter for synchronization [Xia 2016].

**Response Function**

Agents respond to the change of observation via different coefficients $B_i$, which are the response functions of agents. The model does not require only one central observation. There can be local observations by a few neighboring agents and local feedback loops. Agents can have impact only on local observation, i.e. $B_i$ = 0 except for local observations. Agents can also participate and respond to global observations.

In financial markets, $dS$ is people's decision to buy or sell certain securities. $dO$ is the price change. $dE$ is the news outside of price change. When $B_i$>0, the player is a momentum follower; when $B_i$<0, the player is a contrarian.

Agents' power to influence the outcome is denoted as $A_i$, which can vary across agents and time steps.

Let's define total power or wealth as,

$$W(t) = \sum_{i=1}^{N} A_i(t) \tag{3}$$

Change of observation is determined by the aggregated decisions by agents weighted by their power or wealth $A_i$. N is total number of agents.

$$dO(t) = \sum_{i=1}^{N} (A_i/W) * dS_i(t) \tag{4}$$

Agent's action $A_i(t)\ dS_i(t)$ at time $t$ has an instantaneous impact on the outcome $dO(t)$. In Equation (2), agent's decision $dS_i$ at $t$ is influenced by previous observation $dO(t-dt)$ at $t - dt$ and current external force $dE(t)$ at $t$.

$$dS_i(t) = C_i\ (t)\ dE(t) + B_i\ (t)\ dO(t - dt) \tag{5}$$

Hence,

$$dO(t) = \sum_{i=1}^{N} \frac{A_i(t)}{W} * [C_i\ (t)\ dE(t) + B_i\ (t)\ dO(t - dt)\ ] \tag{6}$$

In general $C_i$>0 and agents respond positively to external force $dE$. If the external force persists in the same direction, most agents follow momentum and observation trends in the same direction.

If all agents have $B_i$=0 and $A_i B_i$=0, there is no endogenous reaction, observation simply responds to exogenous signals $dE$.

When there is no external force, $dE(t)$=0, and $B_i$=0, all agents have no action, $dS_i$=0, agents have no impact on observation $dO$. In this case, random noise term $\varepsilon_i$ in equation (1) plays more important role in determining $dO$.

If $dE(t)$=0 and $B_i$ is not zero,

$$dO(t) = \sum_{i=1}^{N} \frac{A_i(t)}{W} * B_i\ (t)\ dO(t - dt) \tag{7}$$

**Useful Energy**

Equation (7) describes an isolated many-body system. The crowd behavior is simply determined by its endogenous properties, or the feedback parameters $A_i$ and $B_i$. We define the useful energy (or effective system power) *D* of a multi-agent system as the weighted aggregate responsiveness of agents to collective observations.

$$D(t) = \sum_{i=1}^{N} A_i(t)B_i\,(t) \tag{8}$$

$$dO(t) = D(t)/W(t) * dO(t-dt) \tag{9}$$

*D*>0 means that the weighted response of the many-body system is to follow the direction of the previous observation change *dO(t-dt)*. *D*<0 means that the weighted response of the system is to reverse the direction of the previous observation change.

If we restrict $-1 \leq B_i \leq 1$, then

$$D(t) \leq W(t) \tag{10}$$

i.e. system's useful energy is always less than total energy. The higher the useful energy, the more impact agents have on the observation, the higher the system output.

**Reward Function**

Wealth reward function was introduced in [Xia 2024] paper. There can be many different reward functional forms. One example of linear reward function is.

$$dA_i(t+dt) = A_i\,(t)\,dS_i\,(t)\,dO(t) \tag{11}$$

From equations (5) and (9) when *dE(t)*=0,

$$dA_i(t+dt) = A_i\,(t) * B_i\,(t)\,dO(t-dt) * D(t)/W(t) * dO(t-dt) \tag{12}$$

Hence, total wealth change

$$\frac{dW(t+dt)}{W(t)} = \frac{\sum_{i=1}^{N} dA_i(t+dt)}{W(t)} = [D(t)/W(t)]^2 * [dO(t-dt)]^2 \tag{13}$$

The growth rate of total wealth (energy) *W* depends on the ratio of useful energy over total energy of the system.

By now, we have introduced system-level macroscopic properties, such as total energy *W* and useful energy *D* linked with agent-level microscopic parameters $A_i$ and $B_i$ . In my previous paper [Xia 2016], order parameter which measures synchronization was also derived from $A_i$ and $B_i$ .

## III. Optimal Order of System

A fundamental question in the study of complex adaptive systems is whether there exists an optimal degree of order. Social systems, biological organisms, economies, organizations, and artificial intelligence networks all rely on interactions among large numbers of agents. Too little coordination may lead to inefficiency, fragmentation, and failure to achieve collective goals. Too much coordination may suppress diversity, reduce adaptability, and increase vulnerability to systemic shocks. The central challenge is therefore to determine how strongly agents should be connected, synchronized, and influenced by one another in order to maximize long-run system performance.

The same tension arises in political and economic systems. Excessive centralization may generate efficiency and rapid mobilization of resources, but it can also increase fragility by concentrating power and suppressing alternative viewpoints. Excessive decentralization may preserve diversity and resilience but reduce collective effectiveness. Acemoglu and Robinson [Acemoglu, & Robinson 2019] describe a similar balance through the concept of “the narrow corridor," in which state capacity is sufficiently strong to maintain order while individual liberty and innovation are preserved. Throughout history, societies have repeatedly oscillated between periods of integration and fragmentation, suggesting that neither extreme is sustainable in the long run.

Financial markets provide another illustration. Periods of strong synchronization among investors can accelerate capital formation, innovation, and economic growth. However, excessive synchronization may also generate bubbles, herding behavior, and systemic crises. Similar phenomena occur in organizations, scientific communities, and online social networks, where highly connected agents can rapidly share information and coordinate actions but may also become susceptible to collective errors and crowd behavior.

These observations suggest that an effective multi-agent system must achieve two objectives simultaneously. First, it must generate productive output through cooperation, specialization, and information sharing. Second, it must preserve sufficient diversity, mobility, and adaptability to withstand shocks and incorporate new information. Productive systems are characterized by high levels of wealth creation, useful energy, innovation, and coordinated action. Resilient systems are characterized by stability, diversity of viewpoints, redistribution of influence, and the ability to adapt to changing conditions.

The concept of optimal order proposed in this paper arises from balancing these competing objectives. We hypothesize that increasing connectivity and synchronization among agents generally enhances collective productivity, but beyond a certain point reduces adaptability and increases systemic fragility. Conversely, increasing independence and diversity enhances resilience but may reduce coordination and productive efficiency. The optimal level of order lies between these extremes, where the system achieves the greatest long-term benefit by balancing growth, stability, and adaptability.

The remainder of this section develops a formal framework for analyzing this tradeoff. We first define a mathematical objective function to optimize by introducing a system-level utility function defined on aggregate power and derive the conditions under which different distributions of agent influence, connectivity, and responsiveness maximize long-run system performance. This framework provides a quantitative foundation for studying optimal order across social, economic, biological, and artificial systems.

**Utility Function**

Objectives of a system can be arbitrary. Hence, objective functions of system optimization depend on subjective choices. There is no single "correct" objective or utility function. As we will see in later discussions, concepts such as order and entropy are also task-dependent and relative to the goals. One's order may be other's disorder.

In thermodynamics, entropy is defined as $S = \frac{W-D}{T} = k\frac{W-D}{W}$, where *T* is temperature and *k* is Boltzmann constant, *W* is total internal energy and *D* is free energy or useful energy, similar to what we had in the previous section. If the objective is to increase free energy (useful energy) *D* while maintain a level of entropy for stability, we can define the objective function as *Max[D*S]* . This simple objective function *D(t)*S(t),* or maximizing the product of useful energy and entropy, is a quadratic function of *D*, $k\left[D - \frac{D^2}{W}\right]$ . We can easily derive that the optimal point occurs at *D=W/2* for this quadratic function. The useful energy *D* is 50% of the total internal energy *W*. This means that the growth rate of total energy is *dW/W=(1/4)** $[\,dO(t - dt)]^2$ in equation (13).

In economics, utility functions provide a formal representation of preferences and objectives. A widely used class of utility functions is the Hyperbolic Absolute Risk Aversion (HARA) family which includes Constant Relative Risk Aversion (CRRA) utility as a special case. CRRA utility takes the form $u(W) = \frac{W^{1-\gamma}}{1-\gamma}$, where *W* denotes wealth and γ denotes the coefficient of relative risk aversion. The parameter γ governs the tradeoff between expected gains and exposure to uncertainty and has become a standard measure of risk preference in macroeconomics, finance, and social welfare analysis. Relative risk aversion is defined as $-\frac{u''(W)W}{u'(W)}$, it equals to $\gamma$.

Quadratic utility approximations have played a particularly important role in portfolio theory. Sharpe's analysis of portfolio choice [Sharpe 2007] and Merton's continuous-time consumption and investment framework [Merton 1969] both rely on mean-variance tradeoffs that can be represented by a quadratic approximation to expected utility, which shows that welfare increases with expected wealth but decreases with wealth volatility. The resulting optimization balances expected growth against risk exposure, a principle that underlies the Sharpe ratio, optimal leverage, and portfolio allocation decisions [Campbell and Viceira 2002].

Utility Function for This Study

Here we introduce utility function of a system as

$$u(W) = W^g \quad (14)$$

Where $g$ is risk appetite parameter, $g>=0$. If $g=0$, maximum risk aversion, $u(W)=1$; if $g<1$, risk aversion; if $g=1$, risk neutral appetite; if $g>1$, risk seeking appetite. Larger $g$ value usually means that there are more agents with positive response function $B_i>0$ .

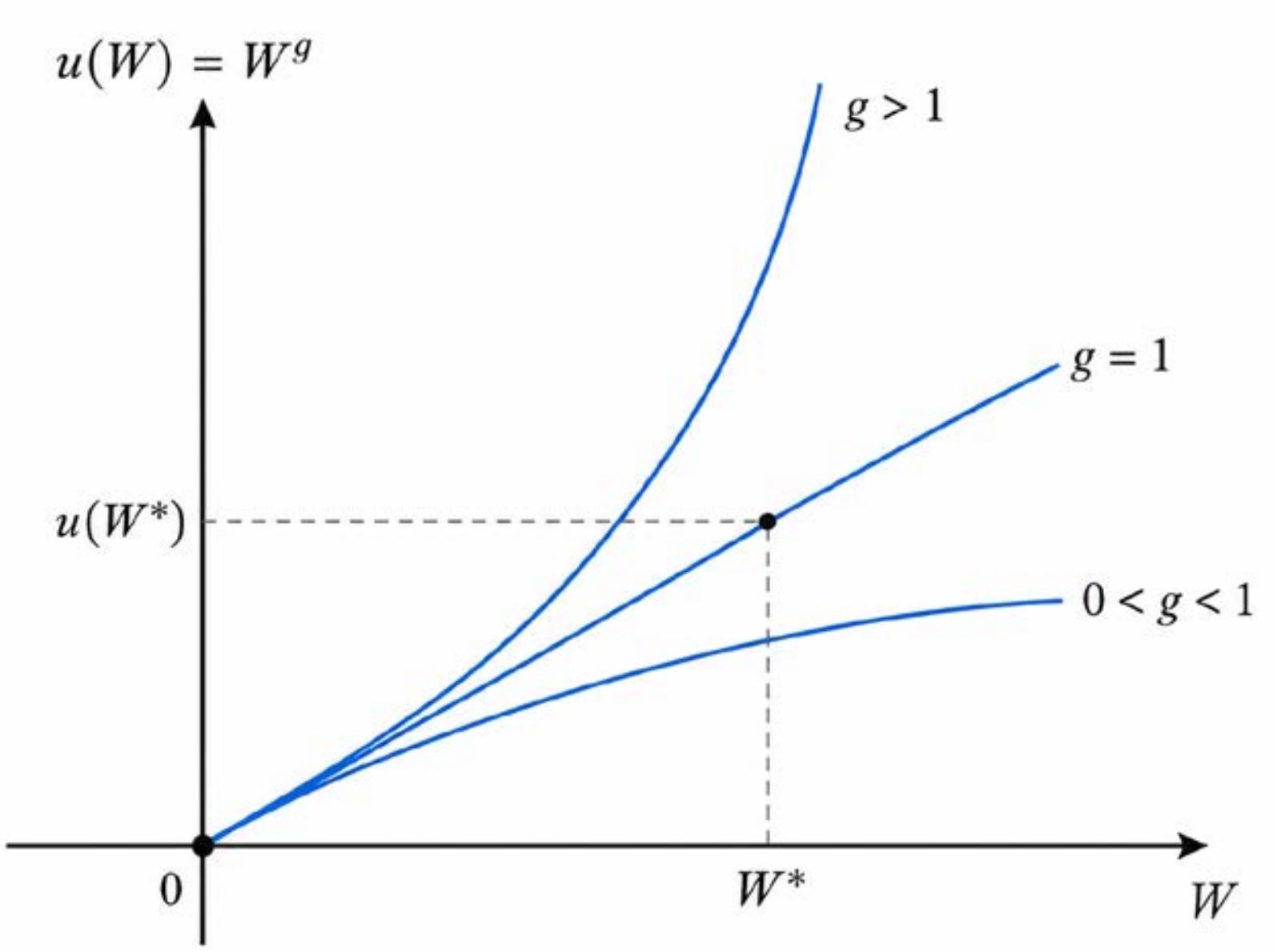

Figure 2. Utility function *u(W)* illustrated for different *g* values.

From relative risk aversion coefficient in CRRA utility function, $\gamma = -\frac{u''(W)W}{u'(W)}$, we can see that risk appetite parameter $g = 1 - \gamma$ .

If $0 < g < 1$, $u(W) = W^g$ is concave, slope of $u(W)$ is steeper at lower $W$, or loss is more painful than gain. For the same amount of $dW$, $u(W)-u(W-dW)>u(W+dW)-u(W)$, or $u'(W-dW)>u'(W+dW)$, $u''(W)<0$.

**Optimization**

For risk aversion systems, $g < 1$. Optimal order happens at balance between bigger growth $R$ and smaller $\sigma_W$. The objective is to maximize expected value of utility function E[$u(W)$]. Using Taylor expansion to the second order, $W(t)$ has a mean $\ddot{W}$ and standard deviation $\sigma_W$ .

$$E[u(W)] \approx u(\ddot{W}) + \frac{1}{2}u''(\ddot{W})\sigma_W{}^2 \quad (15)$$

Since the mean of $W$ is expected to grow at rate of $R$, $\ddot{W} = W_0(1 + R)$ , $W_0$ is the starting value of $W(t)$, $R = \frac{\ddot{W}}{W_0} - 1$ . Since $R = E\left[\frac{dW}{W_0}\right]$ and $dW = W - W_0$ , $E[dW] = \ddot{W} - W_0$ , hence, $\frac{dW}{W_0} = R + \frac{\sigma_W}{W_0}\, dz$ where $dz$ is a Wiener random process.

From equation (15),

$$E[u(W)] \approx \ddot{W}^g + \frac{1}{2}g(g-1)\ddot{W}^{g-2}\sigma_W{}^2$$

$$= W_0{}^g(1 + R)^g + \frac{1}{2}g(g-1)W_0{}^{g-2}(1 + R)^{g-2}\sigma_W{}^2 \quad (16)$$

As we can see, if the system is risk averse, $g$ <1, hence $g - 1$<0. To maximize the expected utility E[$u(W)$], we need to maximize growth rate $R$ while minimizing $\sigma_W{}^2$, which links to the potential maximum loss. For normal distributions, there is less than 16% probability of losing more than $\sigma_W$.

As another example, we can choose a simplified objective function to maximize,

$$G = g * \frac{dW}{W} + (g-1) * \sigma_W{}^2 \quad (17)$$

If $g=0$, maximum risk aversion, $G$=- $\sigma_W{}^2$, the objective is to minimize $\sigma_W$. If $g<1$, risk aversion, the objective is maximize growth of $W$ while minimizing $\sigma_W$. If $g=1$, risk neutral appetite, $G=dW/W$, the objective is to maximize growth of $W$. If $g>1$, risk seeking appetite, the objective is maximize growth of $W$ while also maximizing $\sigma_W$.

In the special case of binary outcome $dO$, only *Sign [dO]* matters. From equation (13), The linear reward function leads to

$$R = E\left[\frac{dW}{W_0}\right] = \frac{D^2}{W_0{}^2} \quad (18)$$

Since $D<=W$, $R<=1$. Uncertainty of $W$, or $\sigma_W$, is coming from unkown factors outside the feedback loop, e.g. unknown external force, random noise in $dS_i$, or simply other external factors.

For general case,

$$dW(t) = \sum_{i=1}^{N} dA_i(t) = W_0 R + \sigma_W dz \quad (19)$$

$$\frac{dA_i(t)}{A_{i0}} = \mu_i + \sigma_i\, dz \quad (20)$$

Where $\mu_i$ is the mean of relative change of $A_i$ , $\sigma_i$ is the standard deviation of relative change of $A_i$ , and $A_{i0}$ is the starting value at previous time step. Hence,

$$R = E\left[\frac{dW}{W_0}\right] = E[\sum_{i=1}^{N} \frac{dA_i}{W_0}] = \frac{1}{W_0}\sum_{i=1}^{N} A_{i0}\ \mu_i \quad (21)$$

Maximum $R$ occurs when every agent has a high growth rate $\mu_i$ .

The variance ${\sigma_W}^2$ depends on both concentration of $A_{i0}$ , and correlation $\rho_{ij}$ of $dA_i$ and $dA_j$ .

$$ {\sigma_W}^2 = \sum_{i=1}^{N} {A_{i0}}^2\, {\sigma_i}^2 + \sum_{i=1}^{N} \sum_{j=1, j\neq i}^{N} \rho_{ij} A_{i0} A_{j0} \sigma_i \sigma_j \quad (22)$$

To minimize ${\sigma_W}^2$ , we need to minimize both the correlation term $\rho_{ij} A_{i0} A_{j0} \sigma_i \sigma_j$ , and the sum term of $\sum_{i=1}^{N} {A_{i0}}^2\, {\sigma_i}^2$ . Correlation $\rho_{ij}$ is determined by unknown factors of $dA_i$. It may or may not be related to $\{B_i\}$ values. In general, the more diverse of $\{B_i\}$, the lower $\rho_{ij}$ .

We can write the sum term in general as variance of *dW/W*,

$$\frac{{\sigma_W}^2}{{W_0}^2} \sim \sum_{i=1}^{N} \frac{{A_{i0}}^2}{{W_0}^2} {\sigma_i}^2$$

In the case of all $\sigma_i = \sigma_1$ ,

$$\frac{{\sigma_W}^2}{{W_0}^2} \sim {\sigma_1}^2 \sum_{i=1}^{N} \frac{{A_{i0}}^2}{{W_0}^2} \quad (23)$$

We can see that when concentration is high, downside risk, reflected in ${\sigma_W}^2$ is high. When correlations between agents are high, ${\sigma_W}^2$ is high. Useful energy in equation (8) $D(t) = \sum_{i=1}^{N} A_i(t) B_i\,(t)$ captures both Ai concentration and Bi correlation.

Concentration can be measured in different ways. Shannon Diversity Index quantifies concentration from probability distribution in a similar formula to information entropy and Gibbs entropy in statistical mechanics. Gini Inequality Index captures dispersion of a distribution. Simpson Concentration Index is defined as $\sum_{i=1}^{N} {p_i}^2$ , where $p_i$ is the probability or weight of category $i$ . The sum term in equation (23) is the same as Simpson Concentration Index measuring distribution of $A_{i0}$.

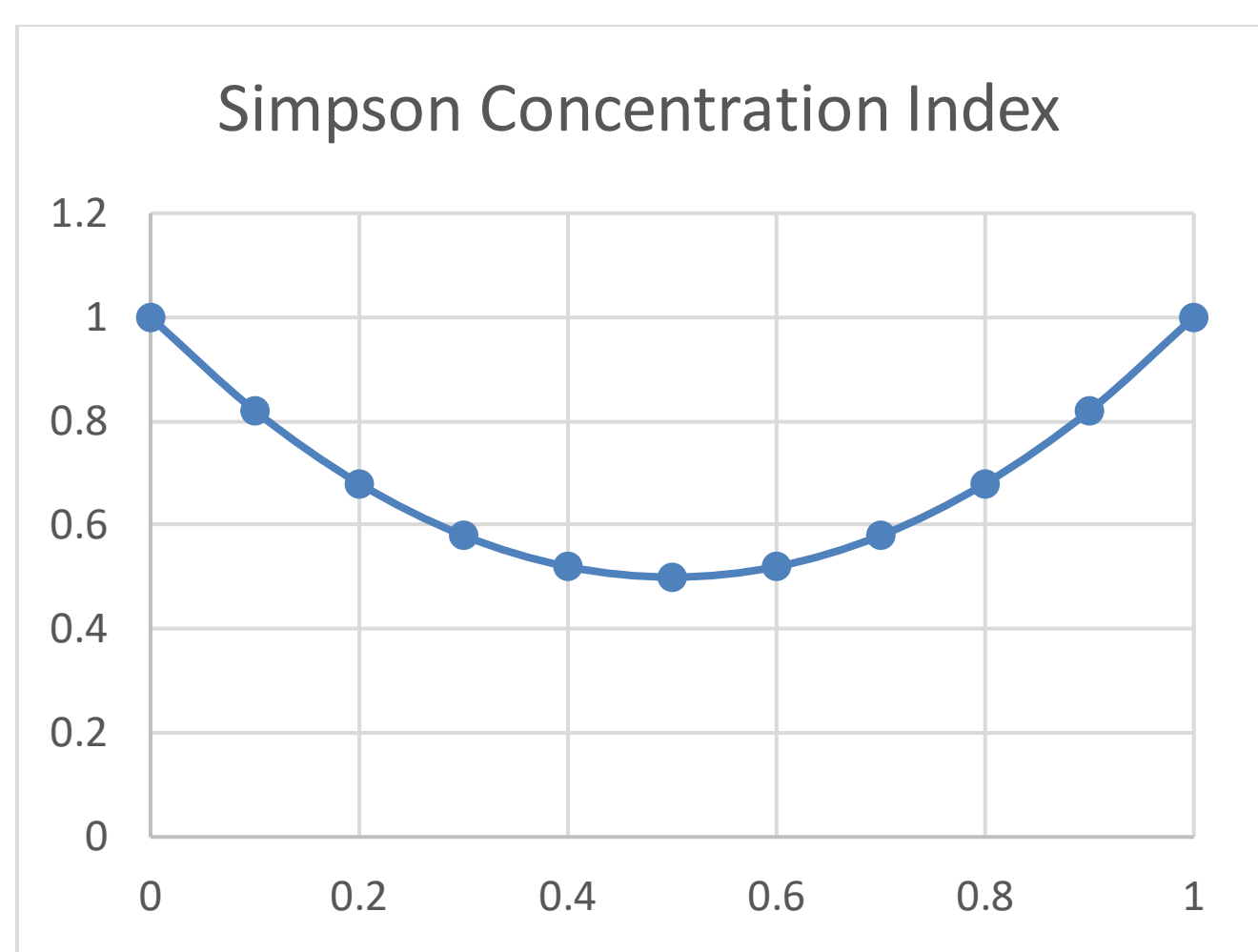

Figure 3. Simpson Concentration Index plotted for special case N=2. Horizontal axis is probability $p_1$ .

For example, if we bet all stake on the outcome of one coin flip, chance of losing all is 50%. But if we bet equal stake on N uncorrelated coin flips, chance of losing all is $(\frac{1}{2})^N$, much lower than 50%. ${\sigma_W}^2 = \frac{{\sigma_1}^2}{N}$ .

**Mobility and Viscosity**

The discussion thus far has focused on two dimensions of social systems: productivity and stability. Productivity reflects the system's ability to generate wealth, useful energy, innovation, and collective output. Stability reflects the system's ability to maintain functionality and avoid large-scale failures. However, these two dimensions alone are insufficient to characterize the quality of a multi-agent system. A third dimension, mobility, plays an equally important role.

Mobility refers to the ability of agents, resources, information, and influence to change positions within a system over time. Mobility cannot be inferred from either growth or volatility. In economic systems, mobility may describe the ability of individuals to move across wealth or income ranks [Carroll & Cohen-Kristiansen 2022]. In organizations, mobility reflects the ability of individuals to assume new roles and responsibilities. In knowledge systems, mobility describes the ability of ideas and beliefs to evolve in response to new evidence. More generally, mobility measures the adaptability and flexibility of a system [Lo & Zhang 2024]. A highly productive but immobile system may become incapable of adaptation or self-repair, while a highly mobile system may fail to accumulate sufficient structure and coordination. The optimal system therefore balances productivity, stability, and mobility.

Mobility (or fluidity) is often modeled as inversely proportional to viscosity. Viscosity represents the resistance of the system to redistribution of power, wealth, influence, or information, which reflects the friction or linkage of agents. One hypothesis of this paper is that viscosity/mobility is related to the strength of feedback interactions $B_i$ among agents. For example, we can hypothesize viscosity as $V = \sum_{i=1}^{N} |B_i|$, mobility is $M=1/V$.

A more generic objective function including mobility to maximize is,

$$G = a_1 * \frac{\mathrm{dW}}{\mathrm{W}} + a_2 * {\sigma_W}^2 + a_3 * M \qquad (24)$$

Where $M$ is mobility, $a_1, a_2, a_3$ are weights related to subjective preferences of growth, risk and adaptability. By maximizing $G$, optimal order can be represented in values of $\{A_i, B_i, C_i\}$. To achieve high growth, useful energy $D$ needs to be big, which requires either high concentration of $A_i$ or high synchronization of $B_i$. To avoid big loss of total power $W$, we need to minimize $\sigma_W$, which requires low concentration of $A_i$ and low correlation of $B_i$. To achieve high mobility of the system, viscosity needs to be low, which requires low absolute values of all $B_i$. The balance of all considerations based on choices of $a_1$, $a_2$ and $a_3$ will arrive at optimal values of $\{A_i, B_i, C_i\}$.

## IV. Order, Entropy, Information, and Useful Energy

Entropy, order, information, and energy are among the most widely used concepts in physics, information theory, and complex system science. However, their meanings often differ across disciplines. In thermodynamics, entropy measures the portion of energy that is unavailable for useful work. In information theory, Shannon entropy measures uncertainty in a probability distribution. Although these concepts are mathematically related, they describe different phenomena and should not be treated as universally equivalent [Georgescu-Roegen 1971].

The present framework focuses on multi-agent systems with dynamic feedback loops, where agents continuously observe, learn, and react to one another. In such systems, the traditional static probability-based definition of entropy is insufficient because system performance depends not only on the distribution of resources among agents but also on the interactions and correlations among them. Two systems may have identical distributions of wealth or influence while exhibiting very different collective behavior due to differences in connectivity, synchronization, and feedback structure.

### What is order?

A central proposition of this paper is that order is not an absolute property. Rather, order is defined relative to a particular task, objective, or receiving system. A configuration that is highly ordered for one purpose may be useless for another. For example, the arrangement of components in a functioning automobile represents an ordered state because it enables transportation. The same collection of components arranged differently contains similar amounts of matter and energy but cannot perform the intended function. Likewise, a sequence of symbols contains useful information only for a receiver capable of interpreting the underlying code or language.

**Entropy**

We therefore distinguish between information entropy, which measures uncertainty of observations or messages, and system entropy, which measures the fraction of total system power that is unavailable for accomplishing a given task. Let $W$ denote total system power and $D$ denote useful power or useful energy. We define system entropy in this framework as

$$S = \frac{W - D}{W} \tag{25}$$

System entropy $S$ is the ratio of useless energy over total energy. $S$ is bounded between 0 and 1. A value of $S=0$ corresponds to a perfectly organized system in which all available power contributes to the task, while $S=1$ corresponds to a system with no useful power, all agents energy are canceled with each other $D(t) = \sum_{\boldsymbol{i=1}}^{\boldsymbol{N}} A_i(t) B_i\,(t) = 0$.

**Information**

From this perspective, information acts as a control signal rather than a source of energy [Toyabe et al 2010]. Information can trigger the release of stored energy, coordinate agent behavior, and alter system dynamics, but its value depends on the receiving system. A password is valuable only to the system it unlocks; a scientific theory is useful only to agents capable of understanding and applying it. Information therefore derives its usefulness from its ability to activate existing structures, resources, or capabilities.

**Useful Energy**

This interpretation naturally connects information, order, and useful energy. Order represents a configuration of agents and resources that enables a system to accomplish a task. Information describes the knowledge required to create or maintain that configuration. Useful energy represents the system's capacity to perform work once the appropriate order has been established. In systems with feedback loops, order is continually created, modified, and destroyed as agents learn from observations and adapt their behavior [Chen 2009].

The framework developed in this paper therefore treats order, entropy, information, and useful energy as task-dependent and system-relative concepts. Rather than viewing entropy solely as a property of probability distributions, we focus on the relationship between collective organization and useful system output.

## V. Discussion and Applications

The central proposition of this paper is that the macroscopic properties of a complex adaptive system emerge from the interaction between agent power and agent response functions. Agent power determines the influence of each agent on collective outcomes, while response functions determine how agents react to observations and to one another. Together, these variables govern the evolution of system order, synchronization, productivity, stability, and adaptability.

From this perspective, the design and management of complex systems can be viewed as the problem of shaping the distribution of power $A_i$ and the response functions $B_i$. Different choices lead to different forms of collective behavior. Excessive concentration of power or excessive responsiveness may increase growth and coordination but also increase fragility and systemic risk. Conversely, weak interactions may increase diversity and adaptability while reducing collective efficiency. The framework therefore provides a common language for studying optimal order across a wide range of domains.

### Large Language Models and Artificial Intelligence

Recent advances in artificial intelligence suggest that large language models (LLMs) can be viewed as complex adaptive systems composed of many interacting computational agents. Individual neurons, attention heads, or expert modules possess limited capabilities, yet collectively generate intelligent behavior through their interactions. From the perspective of the framework developed in this paper, these computational units may be interpreted as agents whose influence is represented by power parameters $A_i$ and whose reactions to system states are governed by response functions $B_i$. Intelligence then emerges from feedback loops among these components rather than from any single agent alone.

This perspective is particularly relevant to recent developments in iterative reasoning architectures, recurrent neural networks, and looped transformer models [Gu & Dao 2023, Saunshi et al 2025], where internal representations are repeatedly refined through feedback. Rather than treating reasoning as a purely feed-forward computation, the framework suggests that reasoning can be understood as a process of collective coordination among competing internal hypotheses. Through repeated interactions and feedback, the system gradually converges toward coherent interpretations and decisions. Power-law behavior occurs in model convergence after scaling up [Kaplan et al 2020, Wei et al 2022], similar to what we applied this model to wealth distribution [Xia 2024].

The theory further suggests that the performance of an AI system depends on achieving an appropriate balance between synchronization and diversity. Insufficient coordination may lead to fragmented or inconsistent reasoning, while excessive synchronization may suppress alternative hypotheses and increase the risk of hallucinations, overconfidence, or systematic errors. Similar to the concept of optimal order in social systems, effective reasoning may require an intermediate level of synchronization that balances consensus formation with adaptability.

The framework therefore provides potential novel design principles for future AI architectures. Rather than focusing solely on increasing model size or computational power, AI systems may be improved by explicitly designing and regulating the distribution of computational influence and response functions within the network. It shifts the focus from scaling parameters to designing the structure of feedback and influence. Measures of synchronization, mobility, and useful power may offer new ways to evaluate reasoning quality, robustness, and adaptability. More broadly, the framework suggests that intelligence in biological brains, social organizations, and artificial neural networks may arise from the same underlying principle: the emergence of optimal order through feedback interactions among heterogeneous agents.

Biological Neural Systems

Biological neural systems provide another example of large-scale adaptive networks. Neurons possess heterogeneous connection strengths and response sensitivities that evolve through learning [Friston 2010]. Memories are encoded through the structure of these connections, while short-term cognition emerges from recurrent feedback loops among active neurons.

Within the present framework, neural activity can be interpreted as a dynamic process in which observations continuously modify response functions and influence distributions. Learning corresponds to changes in connection strength, while synchronization and desynchronization influence cognition, memory formation, and behavior. Phenomena such as addiction may be interpreted as pathological positive feedback loops that excessively amplify specific response pathways.

Autonomous Multi-Agent Systems

The framework is also applicable to decentralized systems of autonomous agents, including robotic swarms, drone networks, and distributed AI systems. Such systems derive robustness not from the strength of individual components but from the adaptability of the collective. Unlike centralized architectures, distributed systems can continue functioning despite the loss of individual agents.

This observation highlights a recurring theme throughout the paper: optimal order often lies between complete centralization and complete independence. Systems that are sufficiently connected to coordinate but sufficiently decentralized to adapt tend to exhibit the greatest resilience to external shocks.

Public Opinion, Social Networks, and Collective Intelligence

Public opinion systems operate through feedback loops between individual beliefs and aggregate social observations. Social media platforms accelerate these feedback processes by increasing the speed and visibility of information transmission. Strong feedback can produce collective intelligence when accurate information spreads efficiently. However, the same mechanisms can also amplify misinformation, polarization, and crowd behavior.

The framework suggests that the quality of collective decision-making depends not only on the diversity of opinions but also on the response functions linking agents. A key challenge is to identify conditions under which minority viewpoints can survive long enough to be evaluated and verified. Future work may explore how different feedback structures influence the balance between crowd wisdom and crowd madness.

Financial Markets and Investment Systems

Financial markets provide a natural laboratory for studying feedback-driven social systems because both wealth and behavior can be directly observed. In this setting, $A_i$ corresponds to the capital controlled by investors, while $B_i$ measures the sensitivity of investors to market observations and the actions of other investors. Highly synchronized markets are characterized by large positive response coefficients, producing trends, momentum, and rapid increases in aggregate wealth. However, the same synchronization can generate speculative bubbles, herding behavior, and systemic crashes.

The framework suggests that market regimes can be characterized by the distribution of response functions and wealth concentration. Monitoring these quantities may provide indicators of crowd synchronization, systemic fragility, and bubble formation. An important area for future research is the empirical estimation of $A_i$ and $B_i$ from market data and the development of early-warning indicators of instability.

Economies, Institutions, and Public Policy

At the national level, governments, firms, and households form a large-scale feedback network. Aggregate output, productivity, and innovation depend not only on the amount of capital within the system but also on the way information and incentives propagate among agents. Excessive concentration of power may increase short-term coordination while reducing adaptability and mobility. Excessive decentralization may preserve diversity but weaken collective action.

The framework provides a quantitative interpretation of long-standing debates concerning markets, governments, and institutions. Concepts such as economic freedom, social mobility, and state capacity can be viewed as properties of the underlying power distribution and response structure. Future research may investigate whether measures of wealth concentration, mobility, and synchronization can explain differences in long-term economic performance across countries.

## VI. Conclusions

To summarize, this paper made the following novel points.

The central idea is that the macroscopic behavior of a system emerges from two fundamental microscopic variables: the distribution of agent power $A_i$ and the response functions $B_i$ that determine how agents react to observations and to one another. We connected macroscopic properties, such as total energy $W(t) = \sum_{i=1}^{N} A_i(t)$, useful energy $D(t) = \sum_{i=1}^{N} A_i(t)B_i(t)$, entropy $[W(t) - D(t)]/W(t)$, mobility or viscosity, with agent-level properties, i.e. agent's power $A_i$ and response function $B_i$.

We introduced a utility function of total energy $u(W) = W^g$ for system optimization. Risk appetite parameter $g$ differentiates system biases, $g = 1 - \gamma$, where γ is the relative risk aversion coefficient.

After selecting objective function to maximize expected value of utility function that strikes the balance between growth of total energy and minimizing losses, we can solve for optimal order from $A_i$ and $B_i$ . In general, higher concentration of agents' power and higher synchronization of agents' behaviors lead to higher growth, but also higher vulnerability to losses. Furthermore, we discussed the importance of system mobility and adaptability, also linked with agent response function $B_i$. This leads to a more general objective function to maximize.

More broadly, we propose that order, entropy, information, and useful energy should be viewed as task-dependent and system-relative concepts. A configuration is ordered only relative to a particular objective, and information is useful only relative to a receiving system capable of acting upon it. Defining task goals is important before solving for optimal order. This perspective provides a common conceptual framework for understanding how collective organization emerges and how useful work is generated in systems with feedback loops.

In applications, besides traditional fields of financial markets, portfolio management, economics, and public opinions, we find the model proposed in this paper can provide particular insights in artificial intelligence models. In particular, recent developments in iterative and feedback-based AI architectures suggest that intelligence itself may be understood as an emergent property of systems that maintain an appropriate balance between synchronization and diversity. From this perspective, designing intelligent systems becomes a problem of organizing power distributions and response functions to achieve an optimal level of order.

Several important questions remain open for future research. These include the empirical measurement of agent response functions, the formal characterization of mobility and adaptability, the relationship between synchronization and systemic fragility, and the development of practical methods for identifying optimal order in real-world systems. Addressing these questions may lead to a deeper understanding of how complex systems grow, adapt, learn, and sustain themselves over time.

Ultimately, the framework proposed here suggests that many seemingly different phenomena—from financial bubbles and social movements to biological cognition and artificial intelligence—can be understood as manifestations of a common principle: the emergence of collective behavior through feedback interactions among heterogeneous agents. By monitoring and designing agent power distributions and response functions, it may be possible to better understand, predict, and optimize the behavior of complex adaptive systems.

ACKNOWLEDGEMENTS: The author would like to thank the following people for helpful discussions [To be completed].